\documentclass[graybox]{svmult}

\usepackage{type1cm}        

\usepackage{makeidx} 
\usepackage{graphicx}
\usepackage{multicol}    
\usepackage[bottom]{footmisc}

\usepackage{newtxtext}       
\usepackage{newtxmath}       
\usepackage[labelfont=bf]{caption}

\makeindex   

\begin{document}

\title*{Modeling the spread of COVID-19 pandemic\\ in Morocco\thanks{This 
is a preprint of a paper whose final and definite form is published in
\emph{Analysis of Infectious Disease Problems (Covid-19) and Their Global Impact},
Springer Nature Singapore Pte Ltd.\\ First submitted 18/April/2020; 
revised 22/Sept/2020; accepted 08/Oct/2020.}}

\author{Houssine Zine, El Mehdi Lotfi, Marouane Mahrouf, Adnane Boukhouima,\\ 
Yassine Aqachmar, Khalid Hattaf, Delfim F. M. Torres and Noura Yousfi}

\authorrunning{Zine, Lotfi, Mahrouf, Boukhouima, Aqachmar, Hattaf, Torres and Yousfi} 

\institute{Houssine Zine 
\at Center for Research and Development in Mathematics and Applications (CIDMA),\\ 
Department of Mathematics, University of Aveiro, 3810-193 Aveiro, Portugal,\\
\email{zinehoussine@ua.pt}
\and El Mehdi Lotfi  
\at  Laboratory of Analysis, Modeling and Simulation (LAMS), Faculty of Sciences Ben M'sik,\\ 
Hassan II University of Casablanca, P.B 7955 Sidi Othman, Casablanca, Morocco,\\
\email{lotfiimehdi@gmail.com}
\and Marouane Mahrouf 
\at  Laboratory of Analysis, Modeling and Simulation (LAMS), Faculty of Sciences Ben M'sik,\\ 
Hassan II University of Casablanca, P.B 7955 Sidi Othman, Casablanca, Morocco,\\
\email{marouane.mahrouf@gmail.com}
\and Adnane Boukhouima 
\at  Laboratory of Analysis, Modeling and Simulation (LAMS), Faculty of Sciences Ben M'sik,\\ 
Hassan II University of Casablanca, P.B 7955 Sidi Othman, Casablanca, Morocco,\\
\email{adnaneboukhouima@gmail.com}
\and Yassine Aqachmar  
\at  World Health Organization (WHO), Country Office, Rabat, Morocco, 
\email{aqachmary@who.int}
\and  Khalid Hattaf 
\at  Centre R\'egional des M\'etiers de l'Education et de la Formation (CRMEF), 
P.B 20340 Derb Ghalef, Casablanca, Morocco,  
\email{k.hattaf@yahoo.fr}
\and Delfim F. M. Torres 
\at  Center for Research and Development in Mathematics and Applications (CIDMA),\\ 
Department of Mathematics, University of Aveiro, 3810-193 Aveiro, Portugal, 
\email{delfim@ua.pt}
\and  Noura Yousfi 
\at  Laboratory of Analysis, Modeling and Simulation (LAMS), 
Faculty of Sciences Ben M'sik,\\ 
Hassan II University of Casablanca, 
P.B 7955 Sidi Othman, Casablanca, Morocco,\\ 
\email{nourayousfi.fsb@gmail.com}}

\maketitle


\abstract{Nowadays, coronavirus disease 2019 (COVID-19) poses a great 
threat to public health and economy worldwide. Unfortunately, there is 
yet no effective drug for this disease. For this, several countries have 
adopted multiple preventive interventions to avoid the spread of COVID-19. 
Here, we propose a delayed mathematical model to predict the epidemiological
trend of COVID-19 in Morocco. Parameter estimation and sensitivity analysis 
of the proposed model are rigorously studied. Moreover, numerical simulations 
are presented in order to test the effectiveness of the preventive measures 
and strategies that were imposed by the Moroccan authorities and also help 
policy makers and public health administration to develop such strategies.}

\medskip



\section{Introduction}
\label{sec:1}

Coronavirus disease 2019 (COVID-19) is an infectious disease that appeared 
in China at the end of 2019. It is caused by a new type of virus belonging 
to the coronaviruses family and recently named \emph{severe acute respiratory
syndrome coronavirus 2} (SARS-CoV-2) \cite{ICTW}. On March 11, 2020, COVID-19 
was reclassified as a pandemic by the World Health Organization (WHO).
The disease spreads rapidly from country to country, causing enormous economic 
damage and many deaths worldwide. The first case of COVID-19 in Morocco 
was confirmed on March 2, 2020 in city of Casablanca. It involved a Moroccan 
expatriate residing in Italy and who came from Italy on February 27, 2020.
As of April 17, 2020, the confirmed cases reached 2564 and the number 
of recoveries reached 281 with a total number of 135 deaths \cite{MHM}.

Moroccan authorities have implemented multiple preventive measures 
and strategies to control the spread of disease, such as the closing 
of borders, suspension of schools and universities, closing coffee shops, 
the shut-down of all mosques in the country, etc. Further, Morocco 
has declared a state of health emergency during the period from March 20 
to April 20, 2020, to avoid the spread of COVID-19. During this period, 
movement during the day should be limited to work, shopping, medical care, 
purchasing medicine, medical supplies, and emergency situations only. 
In addition, and from April 6, 2020, the wearing of a mask became 
compulsory for all persons authorized to move.

Mathematical modeling of COVID-19 transmission has attracted the attention 
of many scientists. Tang et al. \cite{Tang} used a 
Susceptible--Exposed--Infectious--Recovered (SEIR) compartmental model 
to estimate the basic reproduction number of COVID-19 transmission 
based on data obtained for the confirmed cases of the disease in mainland China.
Wu et al. \cite{Wu} provided an estimate of the size of the epidemic in Wuhan 
on the basis of the number of cases exported from Wuhan to cities outside 
mainland China by using a SEIR model. In \cite{Kuniya}, Kuniya applied 
the SEIR compartmental model for the prediction of the epidemic peak 
for COVID-19 in Japan by using the real-time data from January 15 
to February 29, 2020. Fanelli and  Piazza \cite{Fanelli} analyzed and 
forecasted COVID-19 spreading in China, Italy and France by using 
a simple Susceptible--Infected--Recovered--Death (SIRD) model. 
The authors of \cite{MR4093642} present a mathematical model 
and study the dynamics of COVID-19 that emerged recently in Wuhan, China.
For a fractional (non-integer order) model see \cite{Atangana}.

In the models cited above, the transmission of the disease was assumed 
to be instantaneous and therefore they are formulated by ordinary differential 
equations (ODEs), without time delays. In this study, we propose a mathematical 
model governed by delay differential equations (DDEs) to predict the epidemiological 
trend of COVID-19 in Morocco and taking into account multiple preventive measures 
and strategies implemented by Moroccan authorities, related to the confinement 
period between March 2 and June 20, 2020, in order to control the spread of disease. 
To do this, Section~\ref{sec:2} deals with the formulation of the model. 
Section~\ref{sec:3} is devoted to parameters estimation and sensitivity analysis.
Forecast of COVID-19 spreading in Morocco is presented in Section~\ref{sec:4}. 
We end with a discussion of the results, in Section~\ref{sec:5}.


\section{Formulation of the model}
\label{sec:2}

Around the world, all the countries that are attacked by the COVID-19 have imposed 
several strategies, with different degrees, to fight against it, namely the reduction 
of some rights by adopting the quarantine method in order to prevent contacts between 
vulnerable and infected individuals, closing the geographical borders of the countries, 
and enforcing the capacity of the sanitary system. Similarly, the Kingdom of Morocco 
quickly followed all of the previous strategies when the pandemic was in its early stages. 

\begin{remark}
The terms ``susceptibility'' and ``vulnerability'' are often used interchangeably 
for populations with disproportionate health burdens \cite{Y01}. 
The distinction between vulnerability and susceptibility marks the difference between 
being intact but fragile--vulnerable and being injured and predisposed to compound 
additional harm--susceptible \cite{Y02}. Here, we refer to ``the potential to contract 
the COVID-19'' as vulnerability, to emphasize the environmental nature of the disease.
\end{remark}

\begin{figure}	
\begin{center}
\includegraphics[scale=1.00]{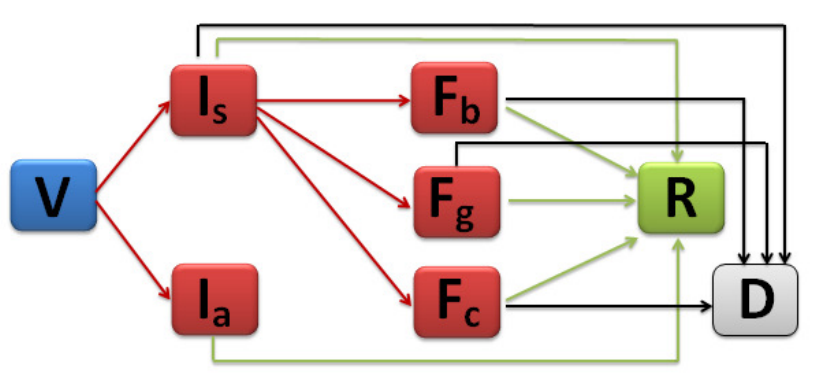}
\captionof{figure}{Schematic diagram of our extended model.}
\label{Flow}
\end{center}
\end{figure}

After the first reported positive case in Morocco, March 2, 2020, the closing of schools 
and universities is done at March 16, 2020; the state of health emergency  (containment) 
is imposed to contain the outbreak from March 20, 2020; and the closure of the borders 
is performed at March 24, 2020. Additionally, the face mask is obligatory used in the 
general population at April 6, 2020. Based on these preventive measures and strategies, 
we model the dynamics  of the  transmission of COVID-19 in Morocco by extending 
the classical SIR model. Precisely, the population is divided into eight classes, denoted by 
$V$, $I_s$, $I_a$, $F_b$, $F_g$, $F_c$, $R$ and $D$, where $V$ represents the vulnerable 
sub-population, which is not infected and has not been infected before, but is susceptible 
to develop the disease if exposed to the virus; $I_s$ is the symptomatic infected 
sub-population, which has not yet been treated, it transmits the disease, and outside 
of proper support it can progress to spontaneous recovery or death; $I_a$ is the 
asymptomatic infected sub-population who is infected but does not transmit the disease, 
it is not known by the health system and progresses spontaneously to recovery; 
$F_b$, $F_g$ and $F_c$ are the patients diagnosed, supported by the Moroccan health 
system and under quarantine, and subdivided into three categories: benign, severe, 
and critical forms, respectively. Finally, $R$ and $D$ are the recovered and died classes.
The schematic diagram of our extended model is illustrated in Figure~\ref{Flow}.

Therefore, the extended model can be governed by the following system of DDEs:
\begin{equation}
\label{Sys1}
\left\{\begin{array}{ll}
\dfrac{dV(t)}{dt}&=-\beta(1-u)V(t)I_{s}(t),\\[0.3cm]
\dfrac{dI_{s}(t)}{dt}&=\beta\epsilon (1-u)V(t-\tau_1)I_{s}(t-\tau_1)
-\left(\mu_s+\eta_{s}+\alpha(\gamma_{b}+\gamma_{g}+\gamma_{c})\right)I_{s}(t),\\[0.3cm]
\dfrac{dI_{a}(t)}{dt}&=\beta(1-\epsilon)(1-u)V(t-\tau_1)I_{s}(t-\tau_1)-\eta_{a}I_{a}(t),\\[0.3cm]
\dfrac{dF_{b}(t)}{dt}&=\alpha\gamma_{b}I_{s}(t-\tau_2)-\big(\mu_b+r_b\big)F_{b}(t),\\[0.3cm]
\dfrac{dF_{g}(t)}{dt}&=\alpha\gamma_{g}I_{s}(t-\tau_2)-\big(\mu_g+r_g\big)F_{g}(t),\\[0.3cm]
\dfrac{dF_{c}(t)}{dt}&=\alpha\gamma_{c}I_{s}(t-\tau_2)-\big(\mu_c+r_c\big)F_{c}(t),\\[0.3cm]
\dfrac{dR(t)}{dt}&=\eta_{a}I_{a}(t)+\eta_{s}I_{s}(t)+r_b F_{b}(t)+r_g F_{g}(t)+r_c F_{c}(t),\\[0.3cm]
\dfrac{dD(t)}{dt}&=\mu_{s}I_s(t)+\mu_{b}F_{b}(t)+\mu_{g}F_{g}(t)+\mu_{c}F_{c}(t),
\end{array}\right.
\end{equation}
where $u$ represents the level of control strategies on the vulnerable population. We adopt the
bilinear incidence rate to describe the infection of the disease and use parameter $\beta$ to denote
the transmission rate. It is reasonable to assume that the infected individuals are subdivided into
individuals with symptoms and others without symptoms, for which we employ the parameter $\epsilon$
to denote the proportion for the symptomatic individuals and $1-\epsilon$ for the asymptomatic ones.
The parameter $\alpha$ measures the efficiency of public health administration for hospitalization.
Diagnosed symptomatic infected population moves to the three forms: benign, severe and critical, by
the rates $\gamma_b$, $\gamma_g$ and $\gamma_c$, respectively. The mean recovery period of these
forms are denoted by  $1/r_b$, $1/r_g$ and $1/r_c$, respectively. The later forms die also with the
rates $\mu_b$, $\mu_g$ and $\mu_c$, respectively. Symptomatic infected population, which is not
diagnosed, moves to the recovery compartment with a rate $\eta_s$ or dies with a rate $\mu_s$. On the
other hand, asymptomatic infected population moves to the recovery compartment with a rate $\eta_a$.
The times delay $\tau_1$ and $\tau_2$ denote the incubation period and the period time needed before
hospitalization, respectively.

For biological reasons, we assume that the initial conditions of system (\ref{Sys1}) satisfy:
\begin{equation*}
\begin{array}{ll}
V(\theta) &=\phi_{1}(\theta)\geq 0,
\quad 
I_s(\theta)=\phi_{2}(\theta)\geq 0, 
\quad 
I_a(\theta)=\phi_{3}(\theta)\geq 0,\\
F_b(\theta) &=\phi_{4}(\theta)\geq 0,
\quad 
F_g(\theta)=\phi_{5}(\theta)\geq 0, 
\quad 
F_c(\theta)=\phi_{6}(\theta)\geq 0,\\
R(\theta)&=\phi_{7}(\theta)\geq 0,
\quad  D(\theta)=\phi_{8}(\theta)\geq 0, 
\quad  \theta\in [-\tau,0],
\end{array}
\end{equation*}
where $\tau=\max\{\tau_1,\tau_2\}$. Let $\mathcal{C}=C([-\tau,0],\mathbb{R}^{8})$ be the Banach
space of continuous functions from the interval $[-\tau,0]$ into  $\mathbb{R}^{8}$, equipped with 
the uniform topology. It follows from the theory of functional differential equations \cite{Hale}
that system (\ref{Sys1}) with initial conditions
$(\phi_{1},\phi_{2},\phi_{3},\phi_{4},\phi_{5},\phi_{6},\phi_{7},\phi_{8})\in \mathcal{C}$
has a unique solution.

On the other hand, the basic reproduction number is an important threshold parameter that
determines the spread of infection when the disease is introduced into the population
\cite{Diekmann}. This number is defined as the expected number of secondary cases produced, 
in a completely susceptible population, by a typical infective individual. By using the next
generation matrix approach \cite{van_den_Driessche}, the basic reproduction number $\mathcal{R}_0$
of system (\ref{Sys1}) is given by
\begin{equation}
\label{18}
\mathcal{R}_0=\rho(FV^{-1})=\dfrac{\beta\epsilon(1-u)}{\eta_{s}+\mu_{s}
+\alpha(\gamma_{b}+\gamma_{g}+\gamma_{c})},
\end{equation}
where $\rho$ is the spectral radius of the next generation matrix $FV^{-1}$ with
$$
F=\begin{pmatrix}
\beta \epsilon (1-u) & 0 \\
0 & 0
\end{pmatrix}  
\quad \text{ and } \quad
V= \begin{pmatrix}
\eta_{s}+\mu_{s}+\alpha(\gamma_{b}+\gamma_{g}+\gamma_{c})& 0\\
0 & \eta_{a}
\end{pmatrix}.
$$


\section{Parameter estimation and sensitivity analysis}
\label{sec:3}

Based on the daily published Moroccan data \cite{url:HIVdata:morocco}, we estimate the values of
some parameters of the model. The proportion of asymptomatic forms can vary from $20.6\%$ 
of infected population to $39.9\%$ \cite{Mizumoto}. Then, $\epsilon \in [0.61, 0.794] $. The
progression rates $ \gamma_b $, $ \gamma_g $ and $ \gamma_c $ from symptomatic infected individuals
to the three forms  are assumed to be $80\%$ of diagnosed cases for benign form, $15\%$ of
diagnosed cases for severe form, and $5\%$ of diagnosed cases for critical form, respectively
\cite{WHO}. The true mortality of COVID-19 will take some time to be fully understood. The data we
have so far indicate that the crude mortality ratio (the number of reported deaths divided by the
reported cases) is between $3$ and $4\%$ \cite{WHO}. As the Moroccan health system is not overloaded at
the moment, it is assumed that deaths mainly come from critical cases with a percentage of $40\%$
for an average period of 13.5 days \cite{WHO}. Since the mortality rate of symptomatic individuals
differs from country to country \cite{Fanelli}, we assume that $1\%$ of symptomatic individuals die
for an average period of 21 days, whereas the recovery rate for asymptomatic cases is $100\%$ and
is the same for severe and benign forms if a proper medical care is taken with an average period of
21 days. We employ a least-square procedure with Poisson noise as in \cite{Kuniya} to estimate the
transmission rate. The incubation period is estimated to be $5.5$ days \cite{WHO1,Stephen} while
the time needed before hospitalization is estimated to be $7.5$ days \cite{Huang,Wang,Haut}. 
The estimation of the above parameters is given in Table~\ref{values1}.
\begin{table}
\captionof{table}{Parameter values for our model (\ref{Sys1}).}
\label{values1}
\begin{center}
\begin{tabular}{ccc}
\hline \hline
Parameter &   Value & Source \\
\hline
$\beta$ & $0.4517$ $(95\%CI, 0.4484-0.455)$ & Estimated\\
$u$ & $0-1$  & Varied \\
$\epsilon$ &   $0.794$ & \cite{Mizumoto} \\
$\gamma_{b}$ &  $0.8$ & \cite{WHO}\\
$\gamma_{g}$ &  $0.15$ & \cite{WHO}\\
$\gamma_{c}$ & $0.05$ & \cite{WHO} \\
$\alpha$ &  0.06 & Assumed\\
$\eta_{a}$ & $1/21$& Calculated\\
$\eta_{s}$ &  $0.8/21$& Calculated\\
$\mu_{s}$ &   $0.01/21$& Calculated\\
$\mu_{b}$ &  $0$& Assumed\\
$\mu_{g}$ &  $0$& Assumed\\
$\mu_{c}$ &  $0.4/13.5$& Calculated\\
$r_{b}$ & $ 1/13.5$& Calculated\\
$r_{g}$ & $1/13.5$& Calculated\\
$r_{c}$ &  $0.6/13.5$& Calculated\\
$ \tau_1 $& $ 5.5$& \cite{WHO1,Stephen}\\
$ \tau_2 $&  $7.5$& \cite{Huang,Wang,Haut}\\
\hline \hline
\end{tabular}
\end{center}
\end{table}

Sensitivity analysis is commonly used to determine the robustness of model predictions to some
parameter values. It is used to discover parameters that have a high impact on $\mathcal{R}_0$ 
and should be targeted by intervention strategies. The main objective of this section is to examine
the sensitivity of the basic reproduction number $\mathcal{R}_0$ with respect to model parameters 
by the so-called \emph{sensitivity index}.

\begin{definition}[\cite{Chitnis,Rodriques}]
\label{sensi}
The normalized forward sensitivity index of a variable $\nu$,
that depends differentially on a parameter $\rho$, is defined as
\begin{equation*}
\label{sensidf}
\Upsilon^{\nu}_\rho := \dfrac{\partial \nu}{\partial \rho}\times\dfrac{\rho}{\nu}.
\end{equation*}
\end{definition}

According to Definition~\ref{sensi}, we derive the normalized forward sensitivity index 
of $\mathcal{R}_0$ with respect to $\beta$, $\epsilon$, $\eta_{s}$, $\mu_{s}$, $\gamma_{b}$,
$\gamma_{g}$, $\gamma_{c}$, and $\alpha$, which is summarized in Table~\ref{sens}.
\begin{table}
\captionof{table}{The normalized forward sensitivity index of $\mathcal{R}_0$.}
\label{sens}
\begin{center}
{\renewcommand{\arraystretch}{2}
\begin{tabular}{ccc} \hline \hline
Parameters &  Sensitivity index of $\mathcal{R}_0$ & Value \\ \hline
$\beta$ & {$\Upsilon^{R_0}_{\beta} \ =\ +1$} & +1\\
$\epsilon$ & {$\Upsilon^{R_0}_\epsilon  
\ =\ +1$}& +1\\
$\eta_{s}$ & {$\Upsilon^{R_0}_{\eta_{s}} 
\ =\ - \dfrac{\eta_{s}}{\eta_{s}+\mu_{s}
+(\gamma_{b}+\gamma_{g}+\gamma_{c})\alpha}$}& -0.3864\\
$\mu_{s}$ & {$\Upsilon^{R_0}_{\mu_{s}} 
\ =\ - \dfrac{\mu_{s}}{\eta_{s}+\mu_{s}
+(\gamma_{b}+\gamma_{g}+\gamma_{c})\alpha}$}& -0.0048\\
$\gamma_{b}$ & {$\Upsilon^{R_0}_{\gamma_{b}} 
\ =\ - \dfrac{\alpha \gamma_{b}}{\eta_{s}+\mu_{s}
+(\gamma_{b}+\gamma_{g}+\gamma_{c})\alpha}$}& -0.487\\
$\gamma_{g}$ & {$\Upsilon^{R_0}_{\gamma_{g}} 
\ =\  - \dfrac{\alpha \gamma_{g}}{\eta_{s}+\mu_{s}
+(\gamma_{b}+\gamma_{g}+\gamma_{c})\alpha}$}& -0.0913\\
$\gamma_{c}$ & {$\Upsilon^{R_0}_{\gamma_{c}} 
\ =\  - \dfrac{\alpha \gamma_{c}}{\eta_{s}+\mu_{s}
+(\gamma_{b}+\gamma_{g}+\gamma_{c})\alpha}$}& -0.0304\\
$\alpha$ & {$\Upsilon^{R_0}_\alpha \ =\ - \dfrac{\alpha (\gamma_{b}
+\gamma_{g}+\gamma_{c})}{\eta_{s}+\mu_{s}+(\gamma_{b}+\gamma_{g}
+\gamma_{c})\alpha}$}& -0.687\\ \hline \hline
\end{tabular}}
\end{center}
\end{table}
As we observe in Table~\ref{sens}, the most sensitive parameters, which have a higher impact on 
$\mathcal{R}_{0}$, are $ \beta $  and $ \epsilon $, since $ \Upsilon^{\mathcal{R}_0}_{\beta} $ 
and $\Upsilon^{\mathcal{R}_0}_{\epsilon} $ are independent of any parameter of system (\ref{Sys1})
with $\Upsilon^{\mathcal{R}_0}_{\beta} =\Upsilon^{\mathcal{R}_0}_\epsilon = +1$. In addition, 
the parameter $\alpha $ has a middle negative impact on $ \mathcal{R}_0 $, while $ \mathcal{R}_0$ 
is slightly impacted by the rest of the parameters.


\section{Prevision of COVID-19 in Morocco}
\label{sec:4}

In this section, we present the forecasts of COVID-19 in Morocco relating with different preventive
measures and strategies implemented by Moroccan authorities on the confinement period between 
March 2 and June 20, 2020. Then the parameter $u$ can be defined as follows:
$$
u=\left\{
\begin{array}{ll}
u_{1}, & \hbox{on $ (\text{March 2},\text{March 10]}$;} \\
u_{2}, & \hbox{on $(\text{March 10},\text{March 20]}$;}\\
u_{3}, & \hbox{on $(\text{March 20},\text{April 6]}$;}\\
u_{4}, & \hbox{after $ \text{April 6}$,}
\end{array}
\right.
$$
where $u_{i}\in (0,1]$, $i=1,2,3,4$, measures the effectiveness of applying the multiple
preventive interventions imposed by Moroccan authorities presented in Table~\ref{inter}.
\begin{table}
\captionof{table}{Summary of non-pharmaceutical interventions considered.}
\label{inter}
\begin{center}
\begin{tabular}{ll}
\hline \hline
\textbf{Policies} &  \textbf{Control values} \\
\hline
Without any intervention measures & $ u=0 $, after March 2 \\ \hline
First set of measures  &  $ u=0.2 $, after March 2 \\ \hline
Second set of measures  &  $ u= 0.2,  \hbox{ on $(\text{March 2},\text{March 10]}$} $ and\\
& $ u= 0.3$, after March 10 \\ \hline
Third set of measures & $ u= 0.2$,  \hbox{ on $ (\text{March 2},\text{March 10]}$}, \\
&  $ u= 0.3$,  \hbox{ on $ (\text{March 10},\text{March 16]}$} and   \\
&  $ u= 0.4$, after March 16\\ \hline
Fourth set of measures & $ u= 0.2$,  \hbox{ on $ (\text{March 2},\text{March 10]}$}, \\
&  $ u= 0.3$,  \hbox{ on $(\text{March 10},\text{March 16]}$} and   \\
& $ u= 0.4$, \hbox{ on $ (\text{March 16},\text{April 6]}$} and   \\
& $ u= 0.8$, \hbox{ after $ \text{April 6}$,} \\ \hline \hline
\end{tabular}
\end{center}
\end{table}
 	
\begin{center}
\begin{figure}	
\includegraphics[scale=0.37]{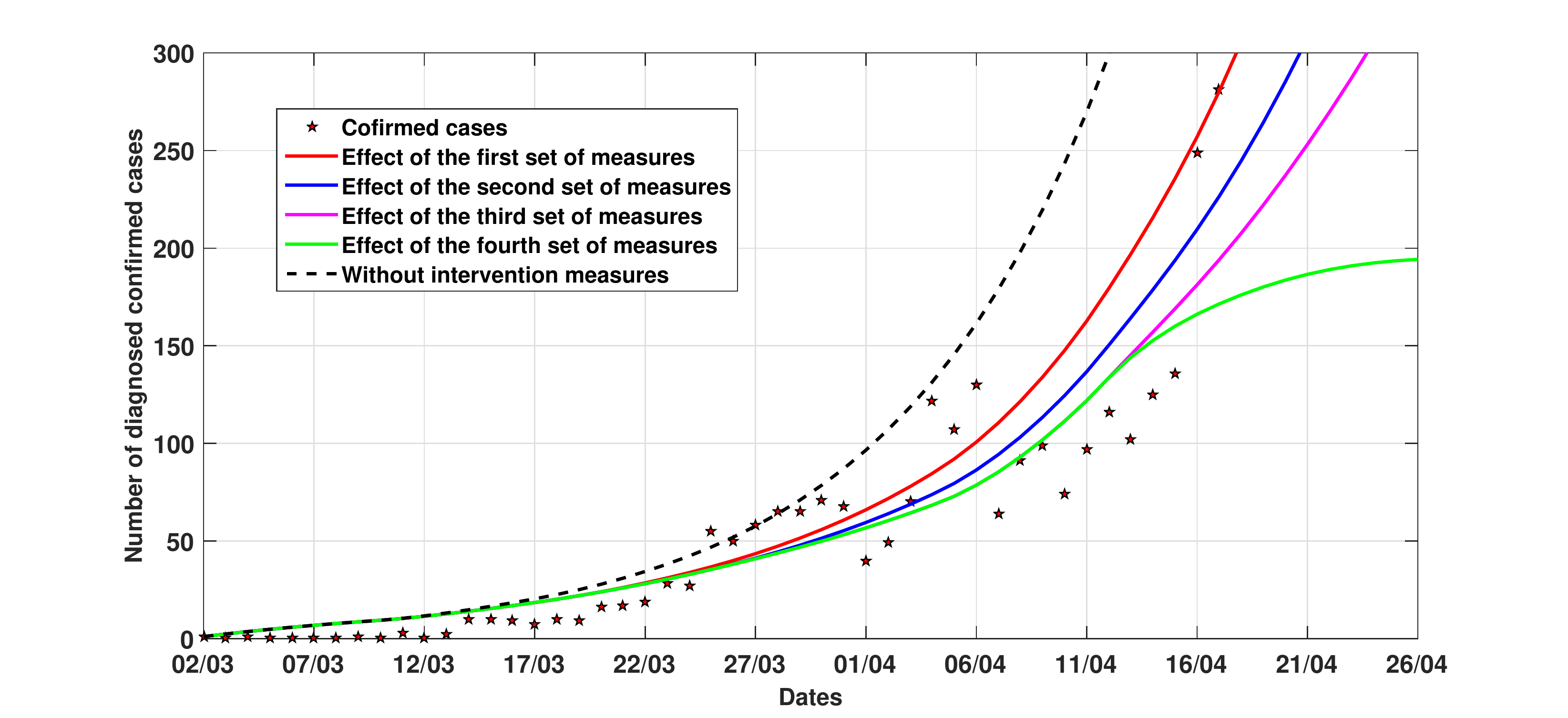}
\captionof{figure}{Comparison of the non-pharmaceutical interventions considered 
and the daily reported cases of COVID-19 in Morocco from March 2 to April 17, 2020.}
\label{Historique}
\end{figure}
\end{center}

To make a better illustration of the different strategies, we test the four decisions 
made at the government level in Figure~\ref{Historique}.
We see in Figure~\ref{Historique} the evolution of the number of diagnosed infected positive
individuals with different sets of measures: low, middle, high, and strict interventions. 
Up to April 15, the curves corresponding to the first three sets of measures increase exponentially,
while the curve corresponding to the fourth set of measures has lost its initial exponential
character and tends to flatten over time. In addition, the last daily reported cases in Morocco from
March 2 to April 17, confirm the biological tendency of our model. Thus, our model is efficient to
describe the spread of COVID-19 in Morocco. However, we note that some clinical data is a little 
far from the values of the model due to certain foci that appeared in some large areas or at the level 
of certain industrial areas.

Next, we give the graphical results related to delays parameters 
to prove their biological importance.
\begin{center}
\begin{figure}	
\includegraphics[scale=0.37]{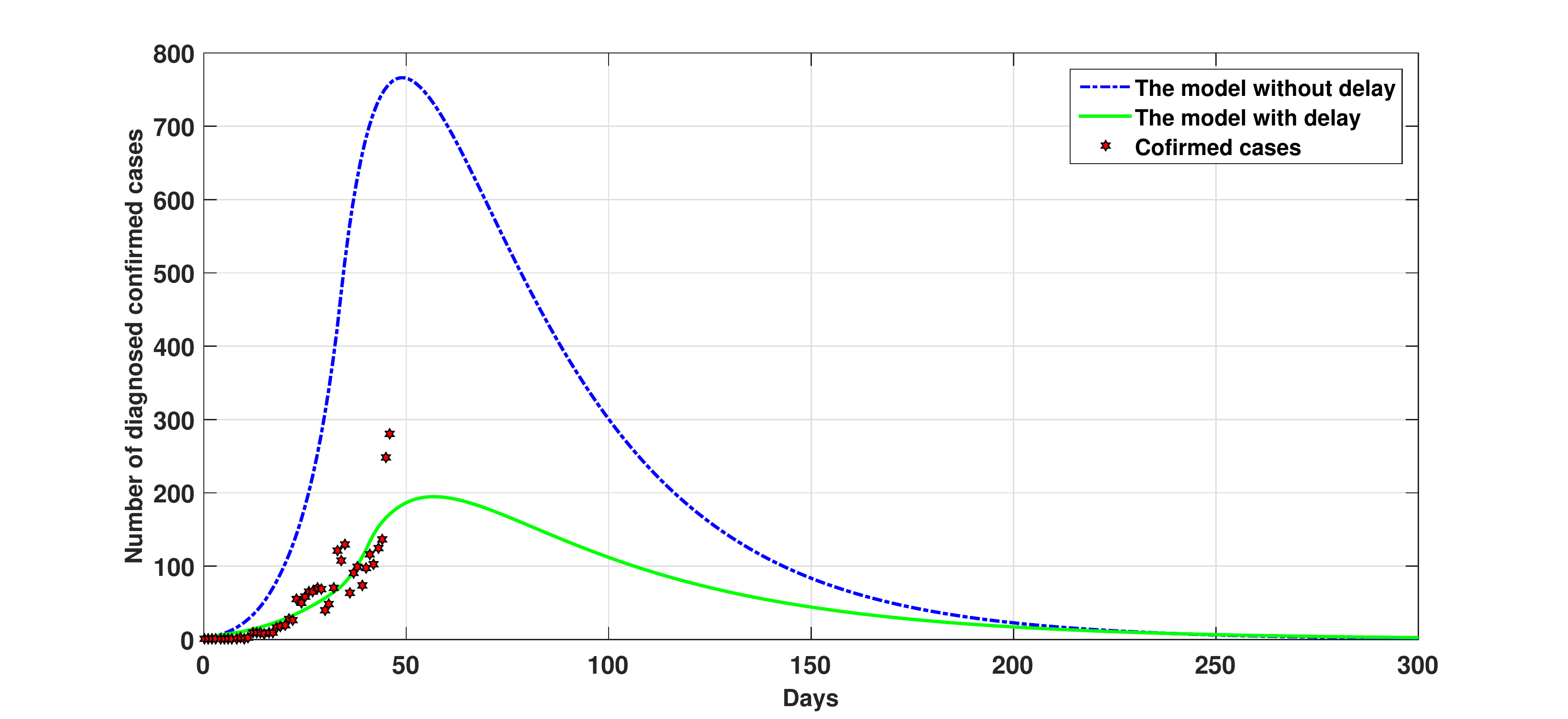}
\captionof{figure}{Effect of delays on the diagnosed confirmed cases. }
\label{Delays}
\end{figure}
\end{center}
We observe in Figure~\ref{Delays} a highly impact of delays on the number of diagnosed positive
cases, thereby the plot of model (\ref{Sys1}) without delays $(\tau_1=\tau_2=0$) is very far 
from the clinical data.


\subsection{Peak prediction}
\label{subsec:4.1}

Now, we indicate the predicted relative impact of the model and especially 
the diagnosed infective individuals with and without interventions applied 
progressively in Morocco.

\begin{center}
\begin{figure}
\includegraphics[scale=0.38]{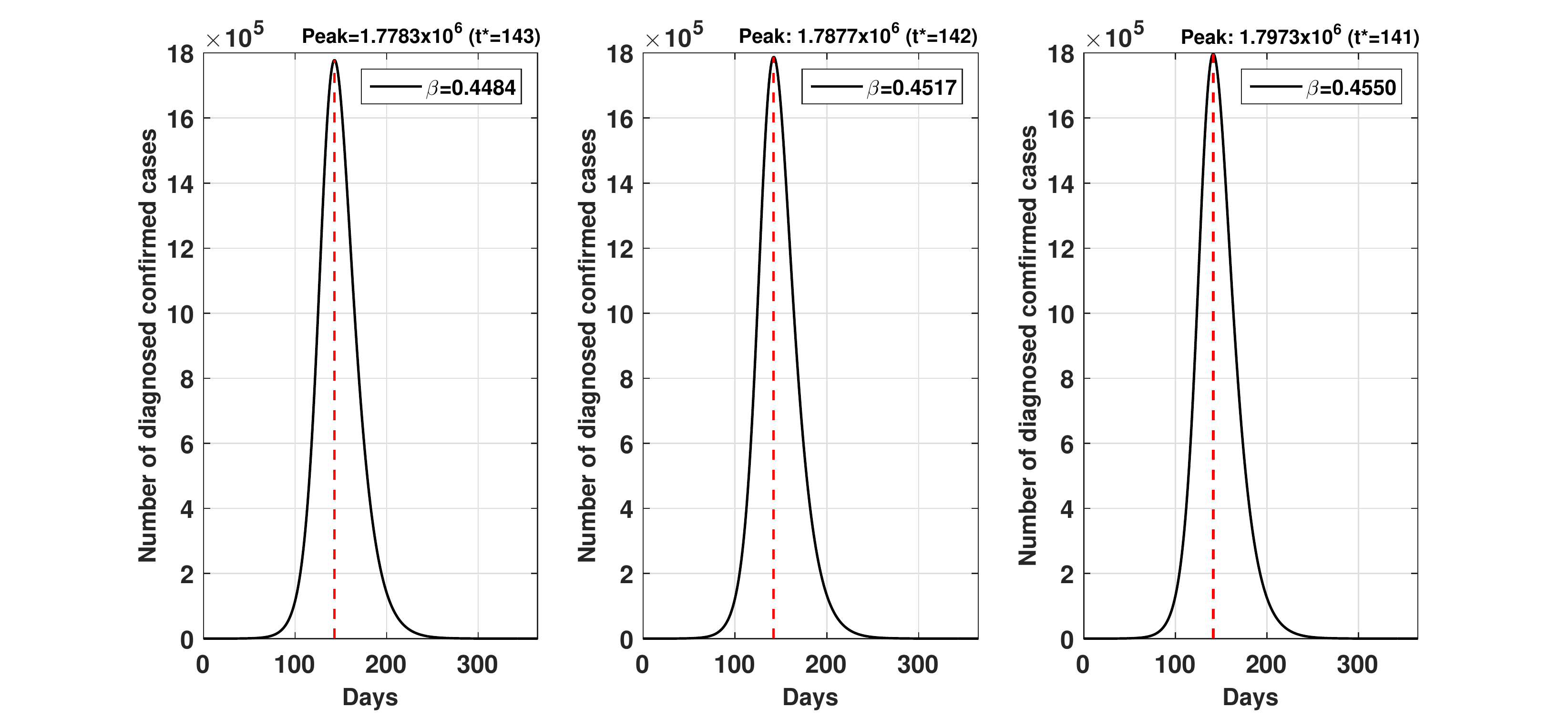}
\captionof{figure}{Time variation of the diagnosed infective individuals without 
any intervention on the Moroccan population with different values 
of $\beta \ (95\%CI, 0.4484-0.455)$.}
\label{Peak}
\end{figure}
\end{center}

\begin{center}
\begin{figure}	
\includegraphics[scale=0.38]{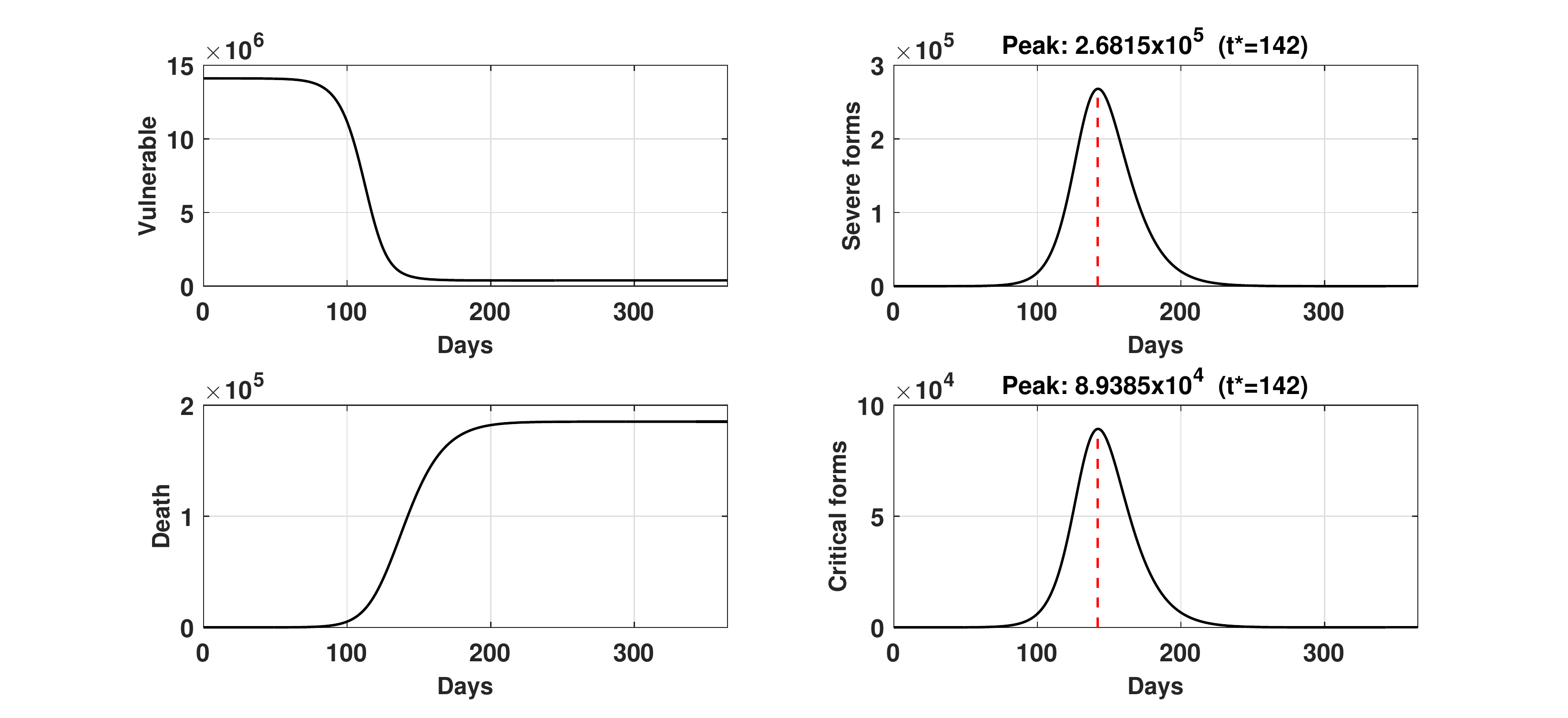}
\captionof{figure}{Time variation of the model with $\beta = 0.4517$ 
and $\mathcal{R}_0=3.6385 $.}
\label{Evol0}
\end{figure}
\end{center}

Before finding the first positive infected case in Morocco, the authorities have begun with 
a suspension of international air lines to and from China, and installed health control 
check-points at the borders but without any interventions into the Moroccan population. 
For this, we simulate model (\ref{Sys1}) in the case $ u=0 $, which is illustrated 
by Figures~\ref{Peak} and \ref{Evol0}.

We remark from Figure~\ref{Peak} that the estimated epidemic peak is 
$t^{*} = 142$ $(95\%CI, 141-143)$, that is, starting from March 2, 
2020 $(t = 0)$, the estimated epidemic peak is July 21, 2020  $(t = 142)$.
In the absence of any government intervention, the disease persists strongly 
and almost all of the vulnerable population will be reached 
by the infection (Figure~\ref{Evol0}).

After the first imported positive infected case, Moroccan authorities began to establish some
preventive interventions between the 2 and the 10th of March, namely isolation of positive cases, 
contact tracing, hygiene measures, prevention measures in workplaces, and ban of mass gathering events. 
For this reason, we have selected in this period $u=0.2$. From March 10 up to March 20, 2020,
additional preventive measures were established: gradual suspension of all international sea, 
ground, and air lines (including with Spain, Italia, Algeria, France, Germany, Netherlands, Belgium, 
and Portugal), closure of coffees, restaurants, cinemas, theaters, party rooms, clubs, sport centers, 
hammams, game rooms and sport fields, closure of mosques, schools and universities, disinfection of public
transportation means, reduction of the carrying capacity of taxis, buses and tramways,
movement/travel restrictions, and containment measures of the general population. These measures
correspond to the choice of the control $u= 0.3$. From March 20 up to April 6, the Moroccan
authority declared a state of emergency with a complete lockdown, night-time curfew, movement
restrictions 24/24, ban of human movements between cities, suspension of railway lines, streets
disinfection, and extensive cleaning and disinfection of port and airport facilities. For this, we
assume that $ u=0.4$. From April 6, the authority decided compulsory wearing of masks in public
spaces, which implies a significant positive influence on the above interventions and an
increase of their efficiency level. In this case, we assume that $u=0.8$. 
Tacking into account all these policies,  we present the following 
Figures~\ref{Peak1} and \ref{Evol1}.
\begin{center}
\begin{figure}	
\includegraphics[scale=0.38]{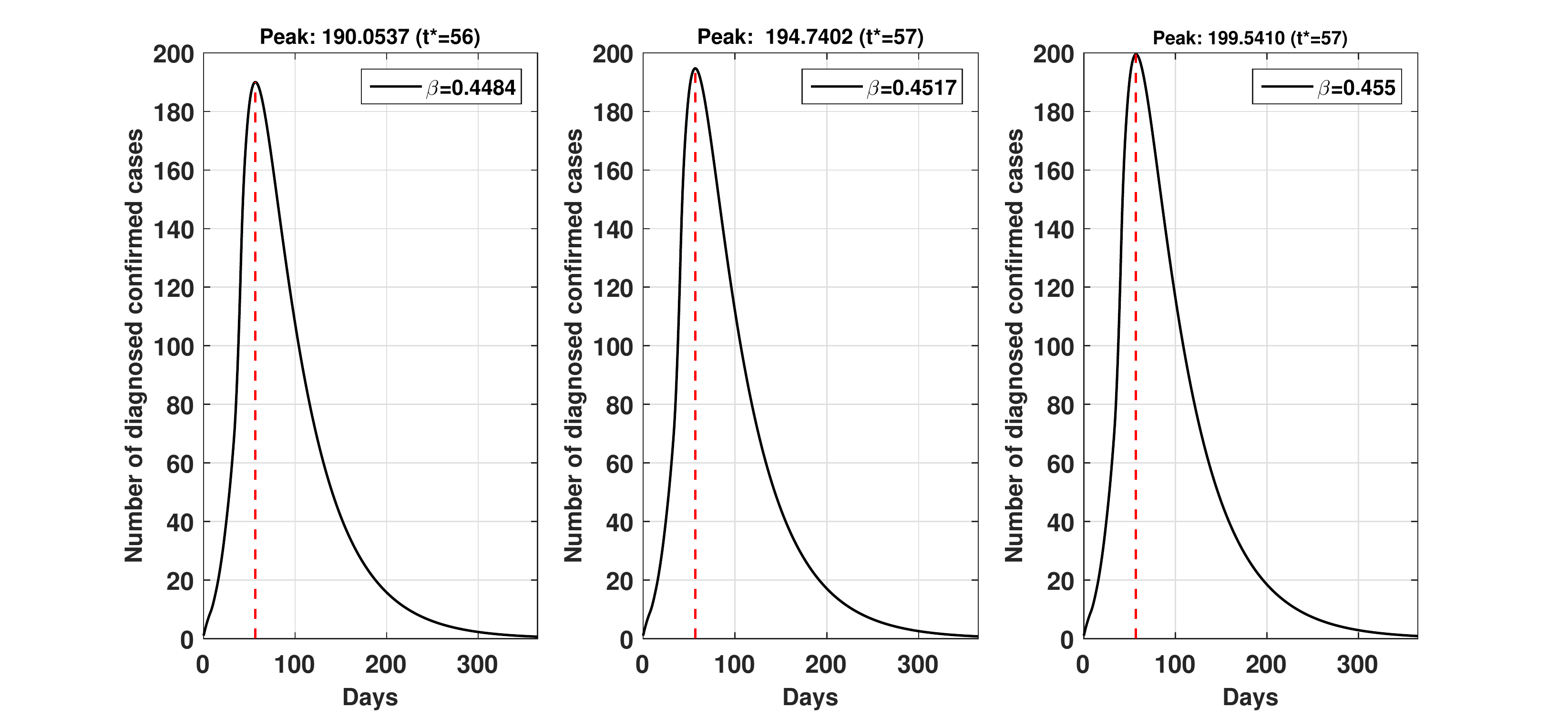}
\captionof{figure}{Time variation of the diagnosed infective individuals 
with hight level respect of measures for different values of 
$\beta$ $(95\%CI, 0.4484-0.455)$.}
\label{Peak1}
\end{figure}
\end{center}
We remark from Figure~\ref{Peak1} that the estimated epidemic peak is 
$t^{*} = 57$ $(95\%CI, 56-57)$, that is, starting from March 2 $(t = 0)$, 
the estimated epidemic peak is April 28, 2020  $(t = 57)$.
\begin{center}
\begin{figure}	
\includegraphics[scale=0.38]{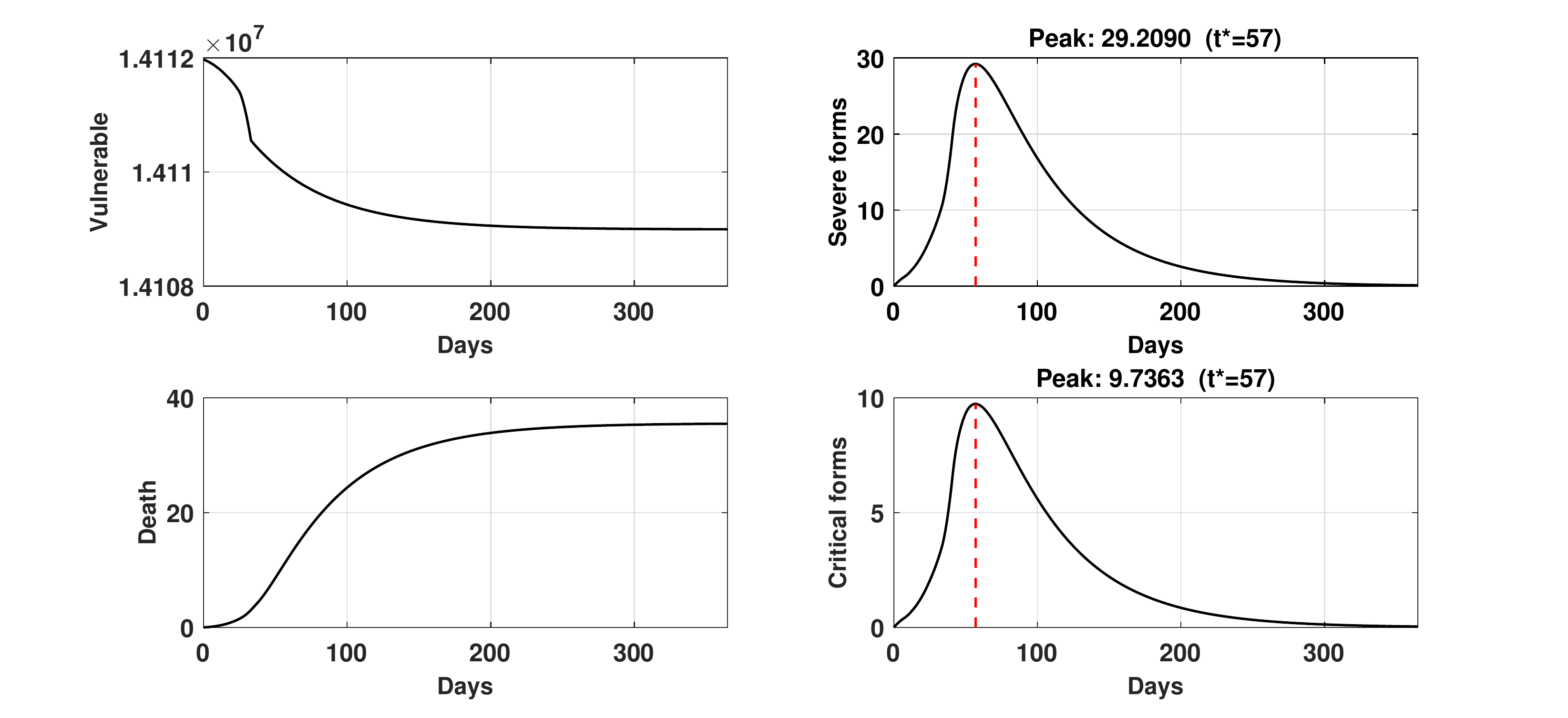}
\captionof{figure}{Time variation of the model with $\beta = 0.4517$ 
and $\mathcal{R}_0= 2.9108$ (March 2--10),
$\mathcal{R}_0=2.5469$ (March 10--20), 
$\mathcal{R}_0=2.1831$ (March 20--April 6), 
$\mathcal{R}_0=0.7277$ (from April 6, 2020).}
\label{Evol1}
\end{figure}
\end{center}
From Figure~\ref{Evol1}, we see that all the measures taken into this second strategy 
have a significant impact on the number of new positive diagnosed cases per day. Compared 
to Figure~\ref{Evol0}, the time required to reach the peak is reduced by 85 days, avoiding globally
an interesting number of new infections and new deaths. Furthermore, the computed basic reproduction
number $\mathcal{R}_0$ is less than $1$, which means the extinction of the disease if the measures
cited above are strictly implemented.


\subsection{Intervention effectiveness}
\label{subsec:4.2}

\begin{center}
\begin{figure}	
\includegraphics[scale=0.38]{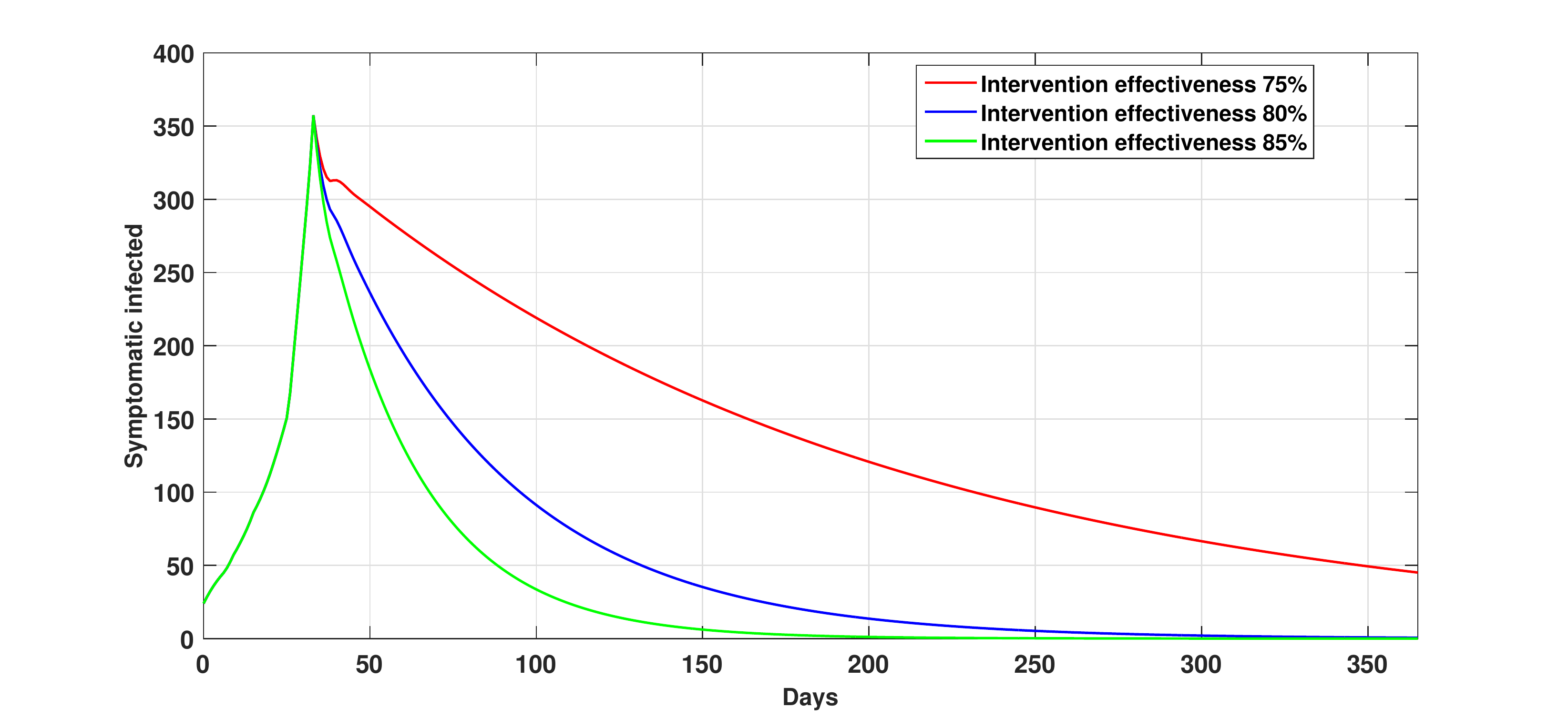}
\captionof{figure}{Evolution of the symptomatic individuals with different effectiveness degrees.}
\label{EffectIs}
\end{figure}
\end{center}

\begin{center}
\begin{figure}	
\includegraphics[scale=0.09]{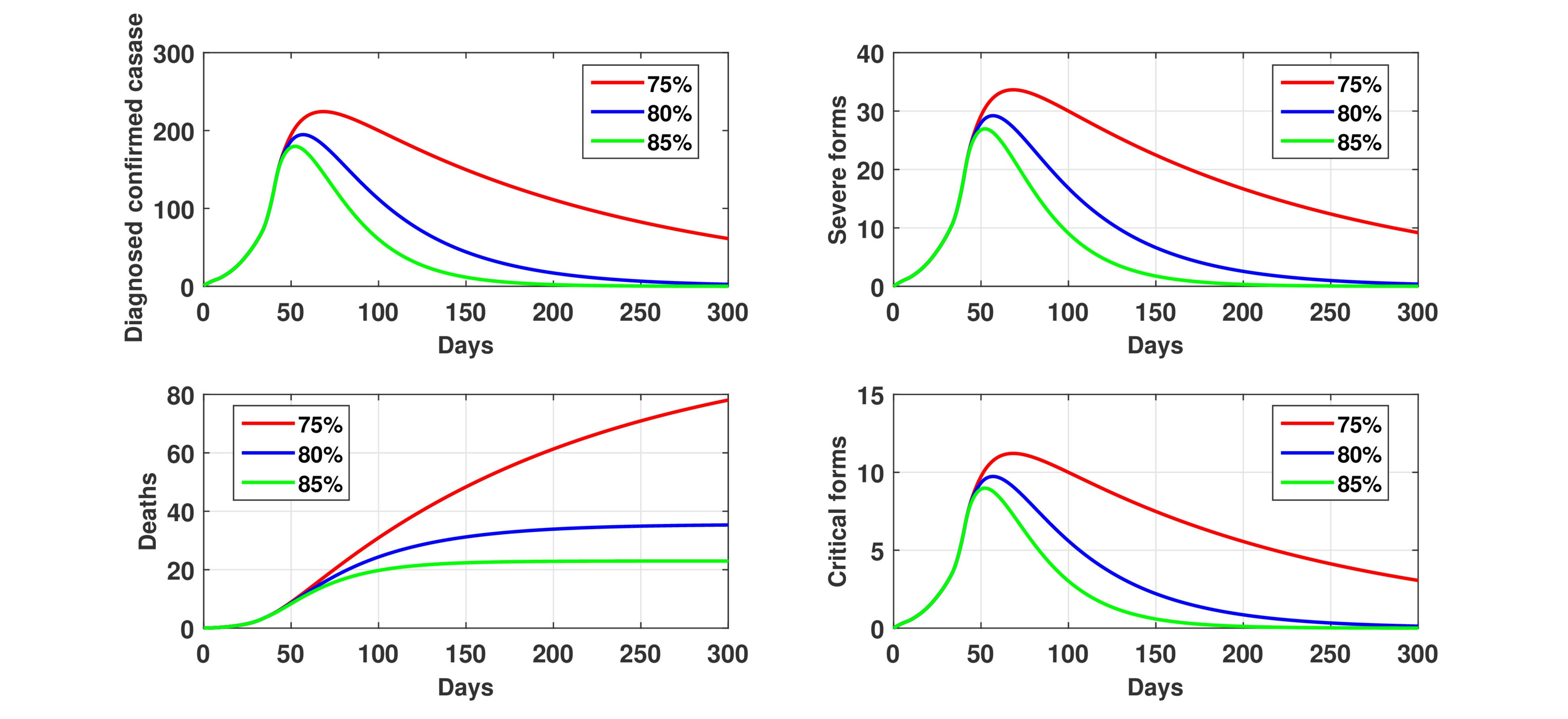}
\captionof{figure}{Evolution of the positive infected diagnosed individuals 
and deaths with different effectiveness degrees.}
\label{Effect}
\end{figure}
\end{center}

\begin{center}
\begin{figure}	
\includegraphics[scale=0.38]{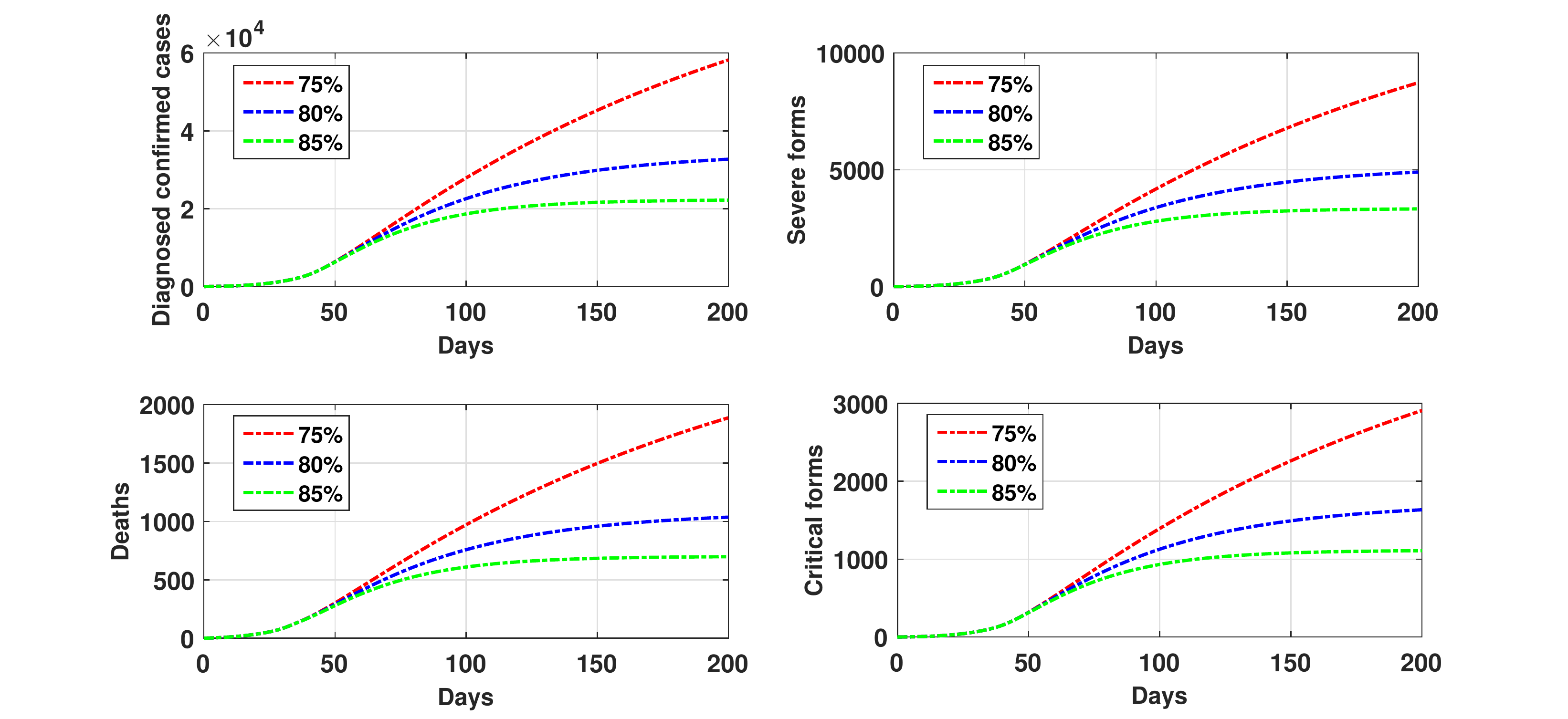}
\captionof{figure}{Cumulative diagnosed cases, severe forms, critical forms, 
and deaths, with different effectiveness degrees.}
\label{Cumulative}
\end{figure}
\end{center}

Here, on one hand, we compare the impact of different degrees of effectiveness on the evolution 
of the number of positive infected diagnosed individuals, symptomatic individuals, and deaths 
(see Figures~\ref{EffectIs} and \ref{Effect}). In addition, we present the cumulative cases in
Figure~\ref{Cumulative} and we summarize it in Table~\ref{Cum}. We remark that the effectiveness
of the policies plays an important role to reduce, or not, the human damage and ensure the eradication
of the illness. However, mitigation measures must be strictly respected to maintain a good level of
control over the spread of the virus.
\begin{table}
\captionof{table}{Cumulative diagnosed cases, severe forms, critical forms, 
and deaths, after 150 days of the start of the pandemic in Morocco.}
\label{Cum}
\begin{center}
\begin{tabular}{lccc}
\hline
\hline
Effectiveness &  $ 75\% $ &  $ 80\% $ &  $ 85\% $ \\
\hline
\hline
Diagnosed &$ 42834$ &$ 29116 $  & $ 21432 $ \\
\hline
Severe forms & $ 6419 $ & $ 4361 $ & $ 3209 $ \\
\hline
Critical forms & $2139$ &$  1453 $  &  $ 1069 $\\
\hline
Deaths & $ 1500 $ & $ 993 $ &$ 661 $\\
\hline
\hline
\end{tabular}
\end{center}
\end{table}
	
\begin{center}
\begin{figure}	
\includegraphics[scale=0.90]{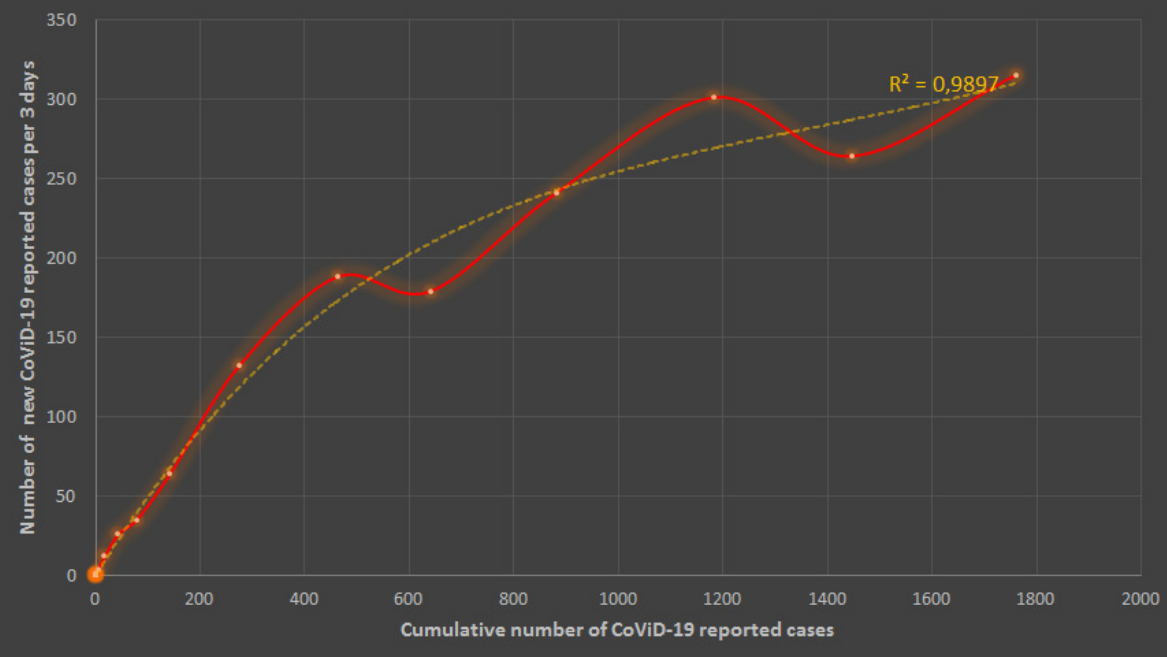}
\captionof{figure}{Trends in the number of new COVID-19 reported cases per three days in Morocco,
compared to the cumulative number of COVID-19 reported cases with correlation coefficient
$R^2=0.9897$.}
\label{yass1}
\end{figure}
\end{center}

\begin{center}
\begin{figure}	
\includegraphics[scale=0.90]{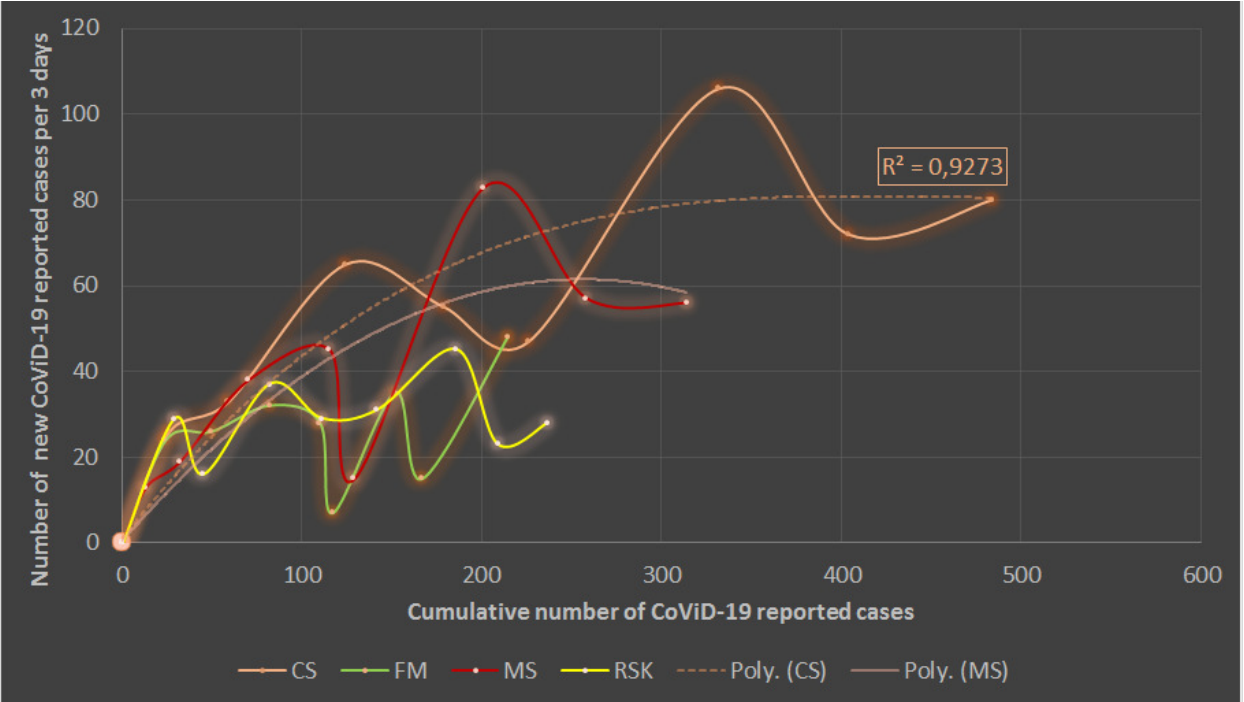}
\captionof{figure}{Trends in the number of new COVID-19 reported cases per three days in Morocco, 
by regions, compared to the cumulative number of COVID-19 reported cases
(CS: Casablanca--Settat; FM: Fes--Meknes; MS: Marrakech--Safi;
RSK: Rabat--Sale--Kenitra).}
\label{yass2}
\end{figure}
\end{center}

\begin{center}
\begin{figure}	
\includegraphics[scale=0.90]{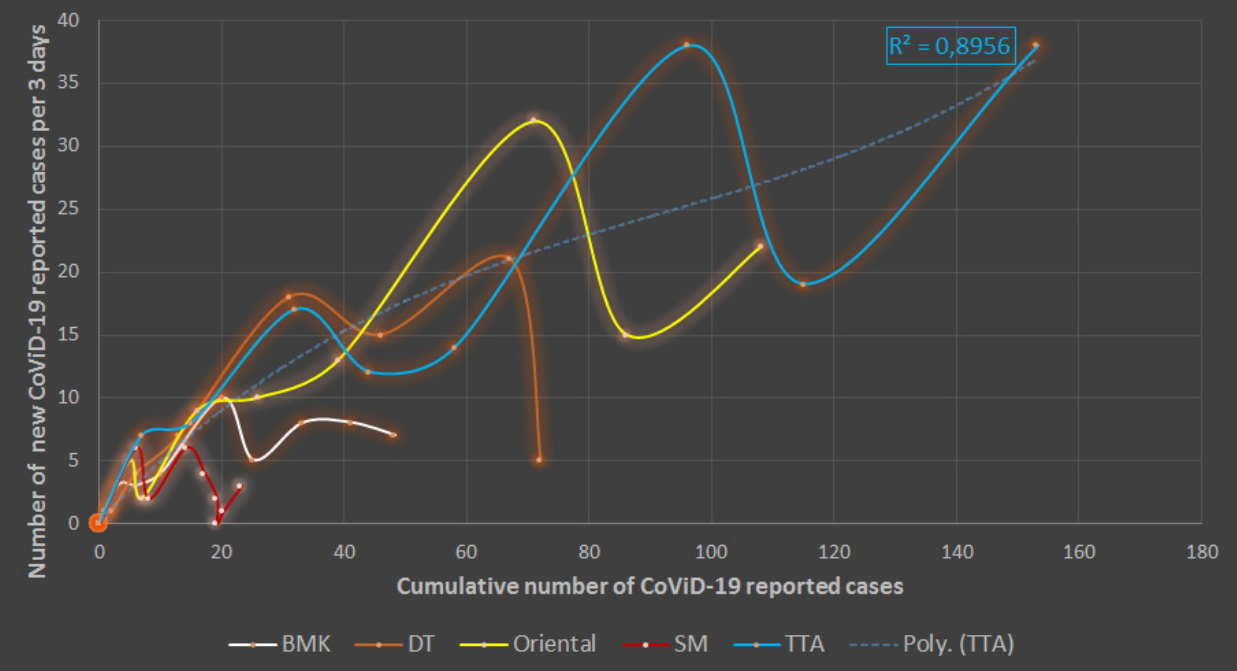}
\captionof{figure}{Trends in the number of new COVID-19 reported cases per three days in Morocco, 
by regions, compared to the cumulative number of COVID-19 reported cases
(BMK: Beni Mellal--Khenifra; DT: Daraa--Tafilalet;
SM: Souss--Massa; TTA: Tetouan--Tangier--Assillah).}
\label{yass3}
\end{figure}
\end{center}

On the other hand, we are carrying out a statistical study on a national scale and we note that
the trend at the beginning was exponential and will undergo a break due to the multiple
interventions of the government, which is globally a good sign (see Figure~\ref{yass1}),
whereas it is needful to pay attention at the evolution of the curves in the different regions 
in Morocco. Since the clinical data of COVID-19 was not available on a daily basis 
at the start of the spread of the epidemic in Morocco, 
we proceeded with a choice of unit of three days. 
We also remark that almost all the regions have a homogeneous tendency with the national 
one, except Tangier--Tetouan--Al Hoceima (TTA), Oriental, Marrakech--Safi (MS), 
and Casablanca--Settat (CS), which show a mitigation of the epidemic that
does not seem very stable (see Figures~\ref{yass2} and \ref{yass3}).


\section{Discussion}
\label{sec:5}

Our work followed several steps. First of all, we have formulated an adequate mathematical model 
to describe the evolution of the COVID-19 disease epidemic in Morocco. This model allows us to have 
an idea on the number of resuscitation beds to prepare for severe forms and intensive care units 
for critical forms. Second, we have estimated the corresponding parameters based on the daily 
Moroccan data. Furthermore, we have applied the least-square-method to determine the confidence 
interval of the transmission rate $ \beta $, which is given by $ (95\%CI, 0.4484-0.455)$. Third, 
we have computed the basic reproduction number $ \mathcal{R}_0 $ with the next-generation-matrix 
method, for which  we have studied the sensitivity analysis in order to examine the robustness 
of the model. We have observed that the transmission rate $ \beta $ and the proportion of 
individuals with symptoms $\epsilon$ are the most sensitive parameters and  have a high impact 
on $\mathcal{R}_0$.  By performing some numerical simulations, we have represented the effect 
of measures taken by the government, step by step, with the control $u$.  In the first period, 
the appropriate basic reproduction number for $u=0.2$ is $\mathcal{R}_0=2.9108$. In the second period, 
$\mathcal{R}_0=2.5469$ for $ u=0.3$. Thirdly, $ \mathcal{R}_0=2.1831$ for $u=0.4$. In the last
period, $\mathcal{R}_0=0.7277$ for $ u=0.8 $. Based on all strategies taken by the Moroccan
authorities, we affirm that the best one is to increase considerably the level of the lockdown
accompanied by the general use of the face masks. In this case, the estimated endemic peak will 
take place around April $28$. Finally, through an analysis of regional data, we have shown that 
the evolution of the pandemic is consistent with the general epidemiological tendency at the
national level. 

We finish by mentioning that the used historic data and the different preventive
measures and strategies implemented by Moroccan authorities and considered in our study, 
are related to the confinement period in Morocco, between March 2 and June 20, 2020, whereas 
the use of historic data and other measures and strategies, linked to the deconfinement phase, 
are left to another research work.


\begin{acknowledgement}
HZ and DFMT were supported by FCT within project UIDB/04106/2020 (CIDMA).
YA brought his expertise in the realization of this work but this does not, 
in any way, reflects an opinion of the World Health Organization (WHO).
The authors would like to express their gratitude to the anonymous reviewers 
for their constructive comments and suggestions, which helped them to enrich 
the work.
\end{acknowledgement}



\end{document}